\documentclass[sigconf]{acmart}
\usepackage{enumitem}
\usepackage{subcaption}
\usepackage[most]{tcolorbox}
\usepackage{booktabs}
\usepackage{tabularx}
\usepackage{multirow, multicol, makecell}

\usepackage[belowskip=2pt,aboveskip=2pt]{caption}

\newtcolorbox{rqsummary}{
    colback=gray!5,
    colframe=gray!40,
    fonttitle=\bfseries,
    coltitle=black,
}

\AtBeginDocument{%
  }

\setcopyright{acmlicensed}
\copyrightyear{2018}
\acmYear{2018}
\acmDOI{XXXXXXX.XXXXXXX}
\acmConference[MSR 2026]{MSR '26: Proceedings of the 23rd International Conference on Mining Software Repositories}{April 2026}{Rio de Janeiro, Brazil}
\acmISBN{978-1-4503-XXXX-X/2018/06}

\begin{document}

\title{Where Do AI Coding Agents Fail? An Empirical Study of Failed Agentic Pull Requests in GitHub}


% \affiliation{%

\author{Ramtin Ehsani}
\orcid{0000-0003-1517-7135}
\affiliation{%
  \institution{Drexel University}
  \city{Philadelphia}
  \state{PA}
  \country{USA}}
\email{ramtin.ehsani@drexel.edu}

\author{Sakshi Pathak}
\affiliation{%
  \institution{Drexel University}
  \city{Philadelphia}
  \state{PA}
  \country{USA}}
\email{sp3856@drexel.edu}

\author{Shriya Rawal}
\affiliation{%
  \institution{Drexel University}
  \city{Philadelphia}
  \state{PA}
  \country{USA}}
\email{sr3728@drexel.edu}

\author{Abdullah Al Mujahid}
\affiliation{%
  \institution{Missouri University of Science and Technology}
  \city{Rolla}
  \state{MO}
  \country{USA}}
\email{amgzc@mst.edu}

\author{Mia Mohammad Imran}
\affiliation{%
  \institution{Missouri University of Science and Technology}
  \city{Rolla}
  \state{MO}
  \country{USA}}
\email{imranm@mst.edu}

\author{Preetha Chatterjee}
\affiliation{%
  \institution{Drexel University}
  \city{Philadelphia}
  \state{PA}
  \country{USA}}
\email{preetha.chatterjee@drexel.edu}

\begin{abstract}
AI coding agents are now submitting pull requests (PRs) to software projects, acting not just as assistants but as autonomous contributors. As these agentic contributions are rapidly increasing across real repositories, little is known about how they behave in practice and why many of them fail to be merged.
In this paper, we conduct a large-scale study of 33k agent-authored PRs made by five coding agents across GitHub. 
\textbf{(RQ1)} We first quantitatively characterize merged and not-merged PRs along four broad dimensions: 1) merge
outcomes across task types, 2) code changes, 3) CI build results, and 4) review dynamics. We observe that tasks related to \textit{documentation}, \textit{CI}, and \textit{build update} achieve the highest merge success, whereas \textit{performance} and \textit{bug-fix} tasks perform the worst. Not-merged PRs tend to involve larger code changes, touch more files, and often do not pass the project's CI/CD pipeline validation. 
\textbf{(RQ2)} To further investigate why some agentic PRs are not merged, we qualitatively analyze 600 PRs to derive a hierarchical taxonomy of rejection patterns.
This analysis complements the quantitative findings in RQ1 by uncovering rejection reasons not captured by quantitative metrics, including \textit{ lack of meaningful reviewer engagement}, \textit{duplicate PRs}, \textit{unwanted feature implementations}, and a\textit{gent misalignment}. Together, our findings highlight key socio-technical and human-AI collaboration factors that are critical to improving the success of future agentic workflows.
\end{abstract}





\maketitle

\vspace{-0.3cm}
\section{Introduction}

AI coding agents such as GitHub Copilot and OpenAI Codex are rapidly becoming active contributors to open-source repositories, often assisting with or directly authoring new pull requests (PRs). Beyond offering inline code suggestions, these tools now generate code changes, respond to reviewer feedback, and participate in the software lifecycle as autonomous agents~\cite{li2025riseaiteammatessoftware,chen2021evaluating,barke2023grounded,yang2024swe,Ehsani_2025}. 
As agent-authored PRs are becoming more prevalent, it is critical to understand how they are evaluated and accepted in practice.

%Because PRs are the primary mechanism through which contributions are evaluated and accepted in modern software projects, they offer a direct window into how agentic contributions succeed or fail. 
%According to GitHub’s 2025 Octoverse report~\cite{staff_octoverse_2025}, nearly half of active open-source projects now include at least one maintainer who uses Copilot, and 72.6\% of developers report improved effectiveness when using it for code review. As a result, agentic PRs are introducing novel forms of human-AI collaboration in real development workflows. With this growing presence, we need to systematically understand how these AI-generated contributions behave in practice.

Prior work shows that PR acceptance depends on factors such as technical correctness, problem scope, and contributor reputation~\cite{lenarduzzi_does_2021,soares2015rejection,9749844}.
%The quality and acceptance of pull requests depend on multiple factors, including the technical soundness of the proposed changes, the importance of the underlying issue, and the reputation or familiarity of the contributor~\cite{lenarduzzi_does_2021}. 
PRs are more likely to be merged when they pass tests and CI pipelines, address high-priority or well-scoped problems, and introduce localized and incremental code changes rather than broad or invasive modifications~\cite{7180094, 8667996, zhao2017impact}. While these factors characterize the success of human-authored PRs, their relevance and applicability to agent-authored PRs are not yet well understood. 
%As AI coding agents increasingly participate in the PR lifecycle, we lack understanding of how often these agentic contributions meet established expectations and under what circumstances they fail. %at which stages of the lifecycle they tend to break down.

Coding agents have been extensively benchmarked across a range of tasks, from code generation~\cite{chen2021codex, sajadi2025llms}, testing~\cite{10.1145/3660771, 10.1145/3597503.3623342, 11205171}, to automated program repair~\cite{jimenez2024swebenchlanguagemodelsresolve, ehsani2025bugfixingbroadercontext, nashid2025characterizingmultihunkpatchesdivergence}. Other studies have analyzed agent-driven code refactoring, reporting that these refactorings tend to be small, localized improvements that produce modest but statistically significant gains in code quality~\cite{horikawa2025agenticrefactoringempiricalstudy,shinn2023reflexion}. More recent work has focused on agent reasoning and execution behaviors, including 
%has examined aspects of agent behavior beyond correctness, such as 
traceability, decision-making, and workflow strategies in complex software engineering tasks~\cite{ceka2025understandingsoftwareengineeringagents, majgaonkar2025understandingcodeagentbehaviour}. 
%However, despite growing interest in understanding how coding agents behave at the system level, very little is known about their behavior when contributing to real software repositories through pull requests. Prior research evaluates agents in isolated tasks, not in the multi-stage, multi-stakeholder lifecycle of real PRs.
While prior work evaluates agents in isolated tasks, we lack a systematic assessment of how agents perform when integrated into real development workflows involving CI validation, code review, and iterative revision.

In this paper, we conduct a large-scale empirical study on agent-authored pull requests using the AIDev-pop dataset~\cite{li2025riseaiteammatessoftware}, which comprises over 33k PRs submitted by five major coding agents across GitHub projects with more than 100 stars. 
We characterize the types of contributions agents attempt, their acceptance rates, reviewer interactions, and, most importantly, where and why their contributions fail.
Specifically, we investigate two RQs:

\noindent\textbf{RQ1: How do merged and not-merged agent-authored PRs differ in task types, code changes, CI outcomes, and review interactions?}
%\textit{Are there any measurable differences in characteristics between merged and unmerged agentic PRs across the PR lifecycle?}
%\textit{How do merged and unmerged agent-authored PRs differ in task types, code changes, CI outcomes, and review interactions?}
%We quantitatively compare structural properties (such as the size and scope of changes, task types, and CI outcomes) and review interactions to determine whether merged and unmerged PRs exhibit distinctive patterns. 
We find that agentic PRs involving \textit{documentation}, \textit{CI}, and \textit{build update} tasks are merged at higher rates, while \textit{performance} and \textit{bug-fix} contributions show the lowest acceptance. Not-merged PRs tend to involve larger code changes, touch more files, receive more reviewer revisions, and frequently fail project CI checks.

\noindent\textbf{RQ2: What patterns lead to agent-authored PRs not being merged in real-world software repositories?}
The most frequent rejection pattern is reviewer abandonment, where agent-authored PRs receive little or no human engagement before being closed. Among PRs that do undergo active review, duplicate PRs, build failures, and unwanted features account for the majority of rejections. 

Overall, our results suggest that agentic PR failures stem from misalignment with repository workflows (e.g., CI/CD failures), developer expectations (e.g., unwanted or incorrect features), and a lack of project coordination (e.g., reviewer abandonment).

\vspace{-0.3cm}
\section{Methodology}

%\noindent\textbf{RQ1: Are there any measurable differences in characteristics between merged and unmerged agentic PRs across the PR lifecycle?}
\noindent\textbf{RQ1: How do merged and not-merged agent-authored PRs differ in task types, code changes, CI outcomes, and review interactions?}

\noindent
We perform a quantitative characterization of agent-authored pull requests along four dimensions: 1) merge outcomes across task types, 2) code changes, 3) CI build results, and 4) review dynamics.
These characteristics are grounded in prior work on factors that influence pull request acceptance~\cite{9749844, 8667996, lenarduzzi_does_2021}.

%\begin{enumerate}[noitemsep, leftmargin=*]
    %\item 
We examine \textbf{merge outcomes across task types}, using the task labels provided in the dataset. These tasks consist of 11 categories: feature, fix, performance, refactoring, style, documentation, test, chore, build, CI, and other~\cite{conv_commits, li2025riseaiteammatessoftware}. This allows us to assess whether certain categories of agent-generated contributions are more or less likely to be merged.
We analyze the magnitude of proposed \textbf{code changes} by measuring a) the total number of added and removed lines of code (\textit{\#LOC Changes}), and b) the number of files modified by each PR (\textit{\#File Changes}). These two characteristics serve as quantitative indicators of the PR’s complexity and potential review burden~\cite{8667996, lenarduzzi_does_2021, 9749844}.
We inspect \textbf{CI build results} for each PR by querying the GitHub API for the status of the last commit in the PR. For every PR, we extract the number of failed check-runs (\textit{\#Failed CI Checks}) and record the overall commit status reported by GitHub (success or failure). This provides a proxy for whether the agent-generated changes break tests, violate linting rules, or fail other repository-specific validation pipelines. This metric allows us to capture automated quality signals that may influence maintainers’ decisions~\cite{8667996, 10298533}.  
We examine \textbf{review interactions} associated with each PR~\cite{8667996}. Specifically, we compute a) the number of \textit{review comments} in a PR (\textit{\#Review Comments}), and b) the number of \textit{review revisions} each PR receives (\textit{\#Review Revisions}). Review revisions are the total number of additions and deletions by the developers during review cycles of the PRs. These measures reflect how much developer attention and iteration an agent-generated PR demands.

Because of the dataset size (> 33k), standard statistical significance testing by itself is not informative because all comparisons might yield statistically significant values even when the differences are negligible~\cite{sullivan_using_2012}. Instead, following best practices in large-scale empirical studies~\cite{7408210, kitchenham_robust_2017}, we rely on effect size measures, using Cliff’s delta ($\delta$) to quantify the magnitude of difference between merged and not-merged PR distributions. To complement effect-size analysis, we use kernel density estimates~\cite{7408210, kitchenham_robust_2017} to visualize distribution shapes. Unlike simple summaries, kernel density plots give a smooth, continuous view of the data’s distribution, making it easier to see shifts and spread in the data~\cite{kitchenham_robust_2017, Thrun_2020}. In addition, we use logistic regression modeling~\cite{kitchenham_robust_2017} to see how effective these characteristics are in predicting the outcome of agent-authored PRs. Together, these analyses allow us to assess whether meaningful differences exist between merged and not-merged agentic PRs.

\noindent\textbf{RQ2: What patterns lead to agent-authored PRs not being merged in real-world software repositories?}

\noindent
We randomly select a subset of 600 rejected PRs for qualitative analysis, stratified across the five coding agents to ensure balanced coverage, and sufficient statistical power to estimate rejection prevalence with 95\% confidence and a ±5\% margin of error~\cite{cochran1977sampling}.

We start with a sample of 100 rejected PRs. 
Following open coding~\cite{azungah_qualitative_2018}, two authors independently label each PR with its primary reason for rejection. During the first round of coding (50 PRs), the annotators identified recurring patterns such as build failures, licensing or contribution policy violations, redundant or unwanted changes, and logical or semantic errors in the proposed code. These patterns were iteratively discussed to derive an initial hierarchical taxonomy spanning three levels: \textsc{Code}, \textsc{Pull Request}, and \textsc{Reviewer} level. 
We assess inter-rater reliability using Cohen’s kappa~\cite{keppaArticle}. After the first round, the agreement was 0.55, indicating moderate alignment. All disagreements were then discussed and resolved through group discussions, during which the taxonomy was refined, and an additional \textsc{Agentic} level was introduced.
The refined taxonomy was applied to a second set of 50 PRs, resulting in a final Cohen’s kappa of 0.91 across all 100 PRs, reflecting strong agreement~\cite{keppaArticle}. Using this taxonomy, the annotators independently labeled 250 more PRs each (500 in total), bringing the size of the manually annotated dataset to 600 PRs. The detailed annotation instructions and codebook are available in the replication package.

Our iterative manual coding resulted in \textit{a hierarchical taxonomy of agentic-PR rejection patterns}, consisting of four high-level categories as described in Table~\ref{tab:failure_modes}. 
1) \textsc{\textbf{Reviewer}} level reflects PRs that close without meaningful developer engagement, often due to inactivity or abandonment.
2) \textsc{\textbf{Pull Request}} level capture cases where the PR is unsuitable for project integration, such as unwanted features, duplicate submissions, or changes submitted to the wrong branch.
3) \textsc{\textbf{Code}} level failures happen when the proposed implementation is incomplete, incorrect, or breaks CI/test pipelines. 
4) \textsc{\textbf{Agentic}} failures capture behaviors such as licensing violations with the project or misalignment with reviewer instructions.
\vspace{-0.3cm}
\section{Results}

\noindent\textbf{RQ1: How do merged and not-merged agent-authored PRs differ in task types, code changes, CI outcomes, and review interactions?}

\noindent
Of the 33,596 agentic pull requests, the majority originate from OpenAI Codex (21,799), exceeding the output of any other agent by more than a factor of four. Copilot and Devin follow with 4,970 and 4,827 PRs, while Cursor contributes 1,541, and Claude Code represents the smallest share with 459 PRs.
Across all agents, 71.48\% of PRs (24,014) are successfully merged. Despite handling the largest volume of contributions, OpenAI Codex achieves the highest merge rate, with 82.59\% (18,004) of its PRs being accepted. Cursor follows with a 65.22\% merge rate (1,005), then Claude Code at 59.04\% (271), and Devin at 53.76\% (2,595). Copilot exhibits the lowest merge rate, with only 43.04\% (2,139) of its PRs being merged.

\begin{figure}[h]
    \centering
    \vspace{-0.35cm}
    \includegraphics[width=\linewidth]{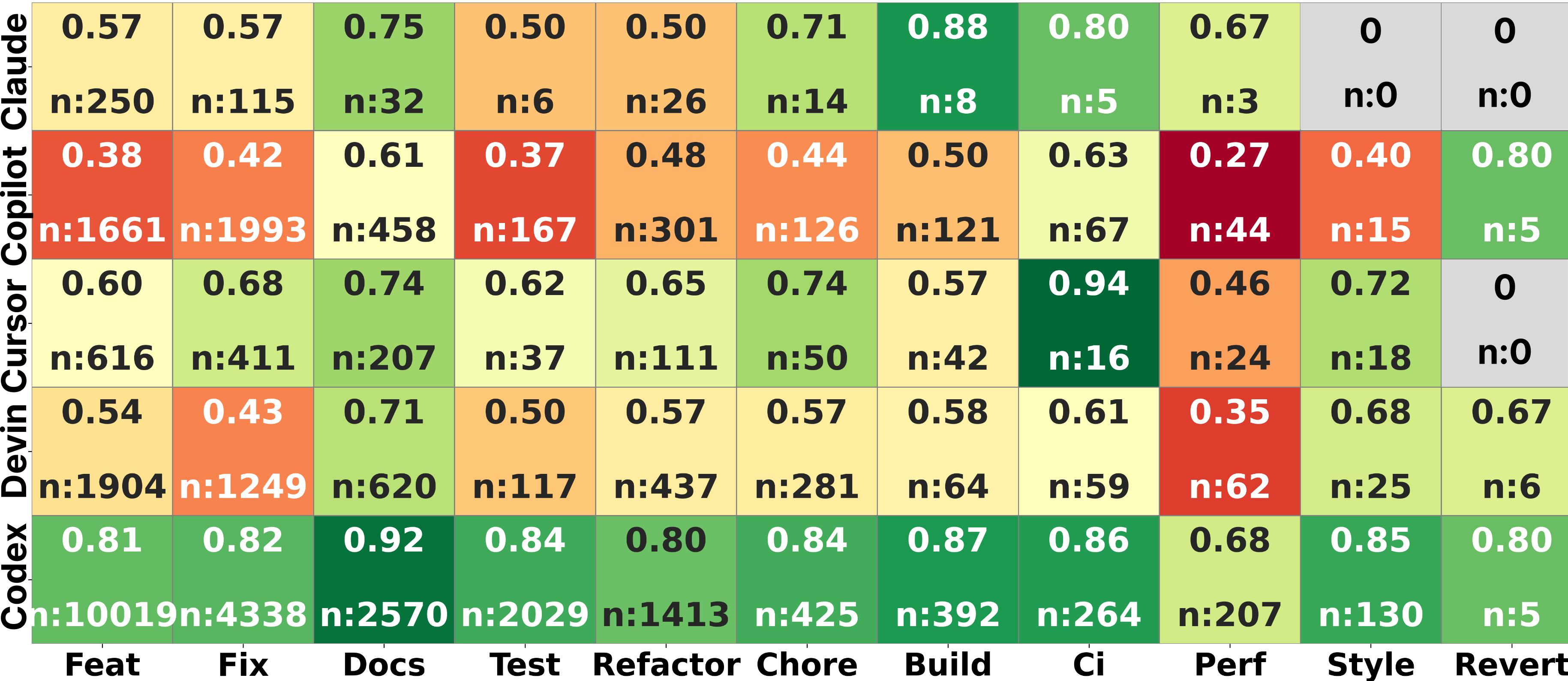}
    \caption{Merge-rate per Task Type Across Agentic PRs.}
    \label{fig:task_types}
    \vspace{-0.45cm}
\end{figure}

Figure~\ref{fig:task_types} shows the merge rates for each agent across all \textbf{task categories}, along with the number of PRs in each cell. The distribution of success varies across task types and agents. Codex shows the highest merge rates overall, exceeding 80\% in categories including \textit{documentations} (0.92), \textit{CI} (0.86), \textit{build} (0.87), \textit{chore} (0.84), \textit{test} (0.84), \textit{fix} (0.82), \textit{feature} (0.81), and \textit{refactoring} (0.80). Its lowest category is \textit{performance} (0.68).
Cursor also performs strongly on maintenance-oriented tasks, with high merge rates in \textit{CI} (0.94), \textit{documentations} (0.74), \textit{chore} (0.74), and \textit{style} (0.72). Its lowest rates occur in \textit{build} (0.57) and \textit{performance} (0.46). Claude Code’s highest merge rates appear in \textit{build} (0.88), \textit{documentations} (0.75), and \textit{CI} (0.57), with lower rates for \textit{test} (0.50) and \textit{refactoring} (0.50). Devin shows moderate values across most categories, with higher merge rates in \textit{documentations} (0.71), \textit{style} (0.68), and \textit{CI} (0.61), and lower rates in \textit{performance} (0.35) and \textit{fix} (0.43). Copilot displays the lowest merge rates among the agents, with its highest values in \textit{CI} (0.63) and \textit{documentations} (0.61), and its lowest in \textit{feature} (0.38), \textit{test} (0.37), and \textit{performance} (0.27).
Across agents, tasks with consistently higher merge rates include \textbf{documentations} (84\%), \textbf{CI} (79\%), and \textbf{build} (74\%).
% each showing multiple cells above 70\%. 
In contrast, \textbf{performance} (55\%) and \textbf{fix} (64\%) display the lowest merge rates overall.
These results indicate that tasks involving documentation, CI, and build changes tend to merge more easily in repositories, whereas categories requiring more complex or subjective changes show lower merge rates.

\begin{figure}[tb]
    \centering
    \begin{subfigure}{0.42\columnwidth}
        \centering
        \includegraphics[width=\linewidth]{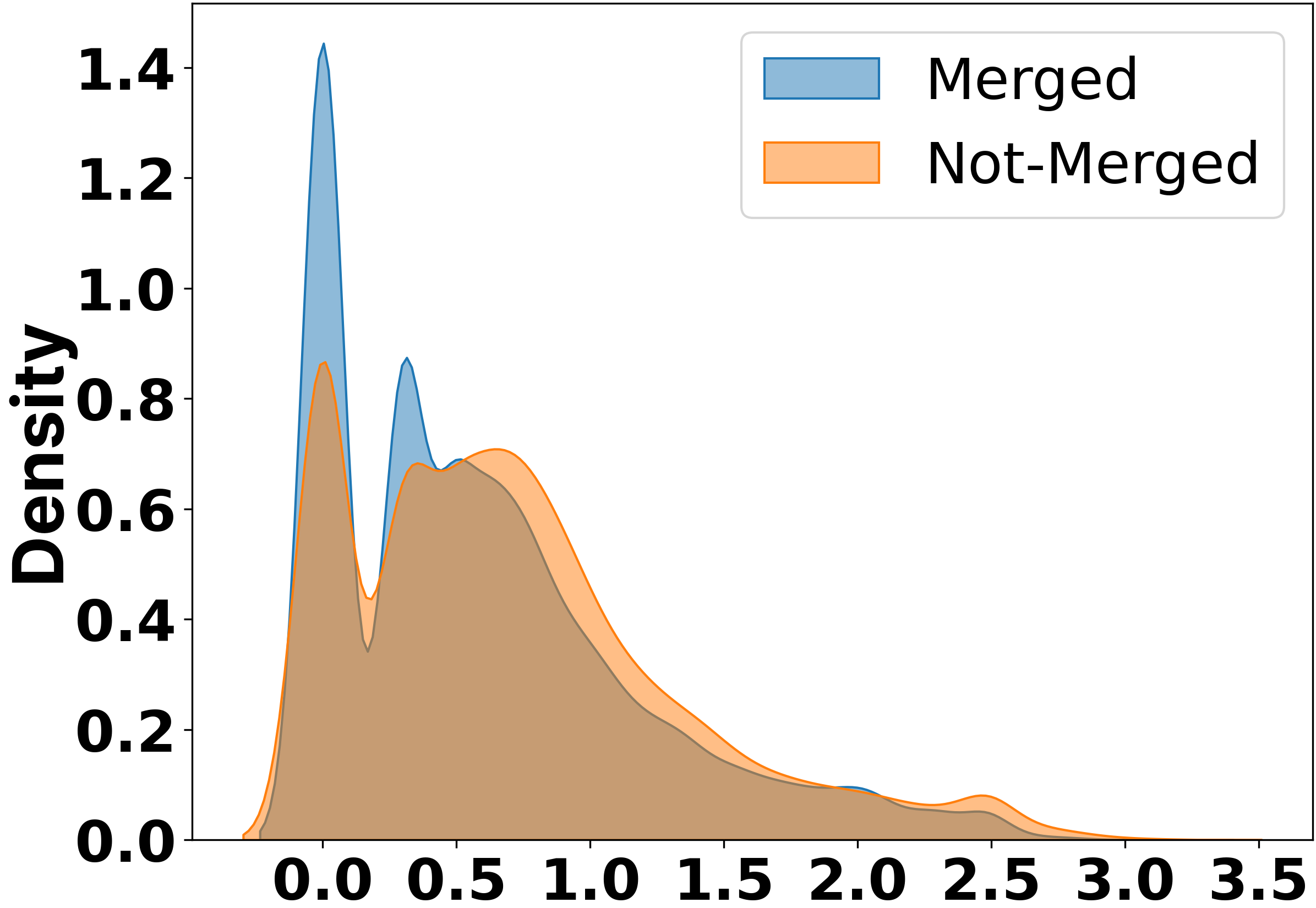}
        \caption{Changed Files (log10)}
        \label{fig:file_changes}
    \end{subfigure}
    \hfill
    \begin{subfigure}{0.4\columnwidth}
        \centering
        \includegraphics[width=\linewidth]{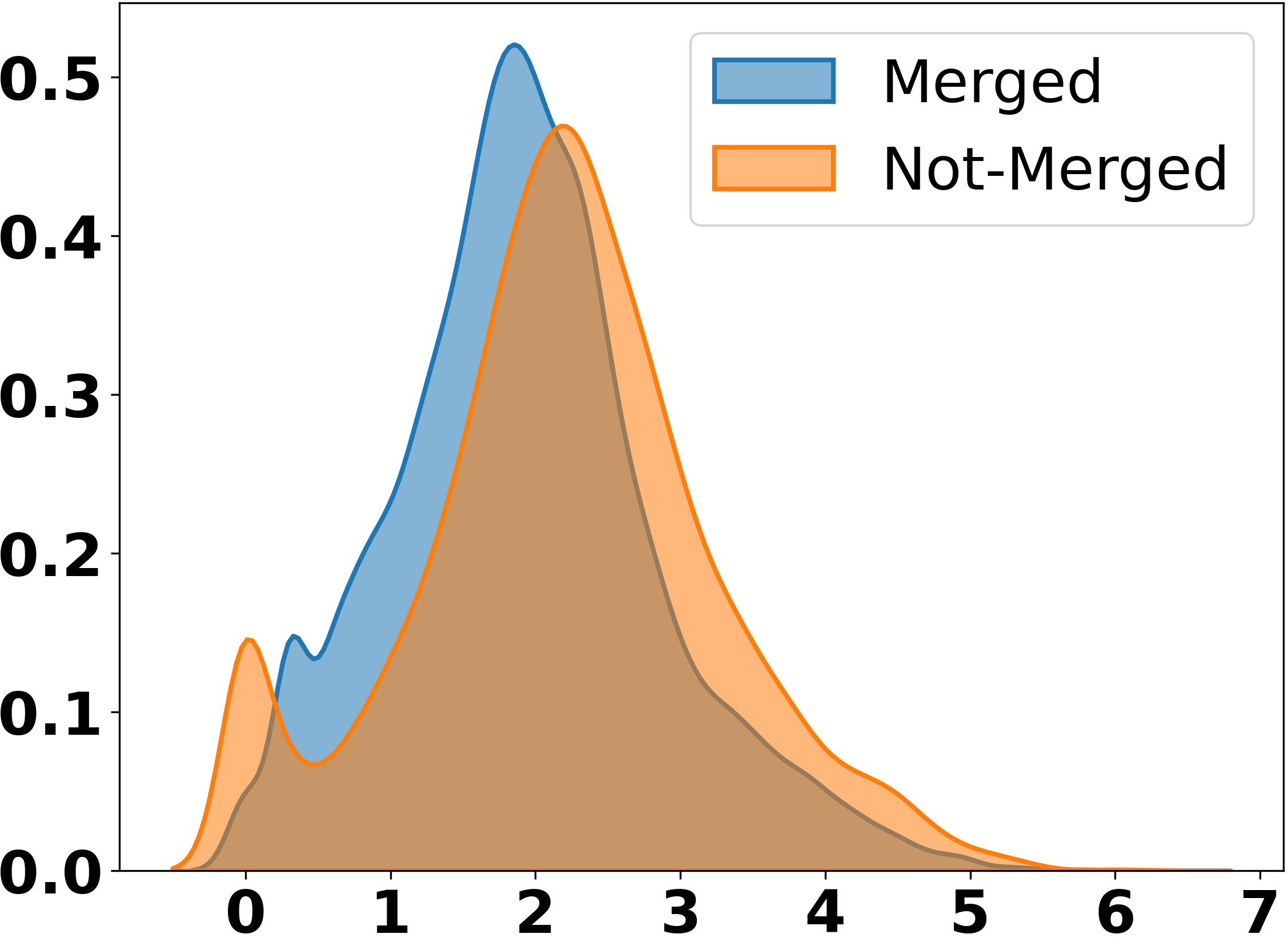}
        \caption{Changed LOC (log10)}
        \label{fig:code_changes}
    \end{subfigure}
    \begin{subfigure}{0.4\columnwidth}
        \centering
        \includegraphics[width=\linewidth]{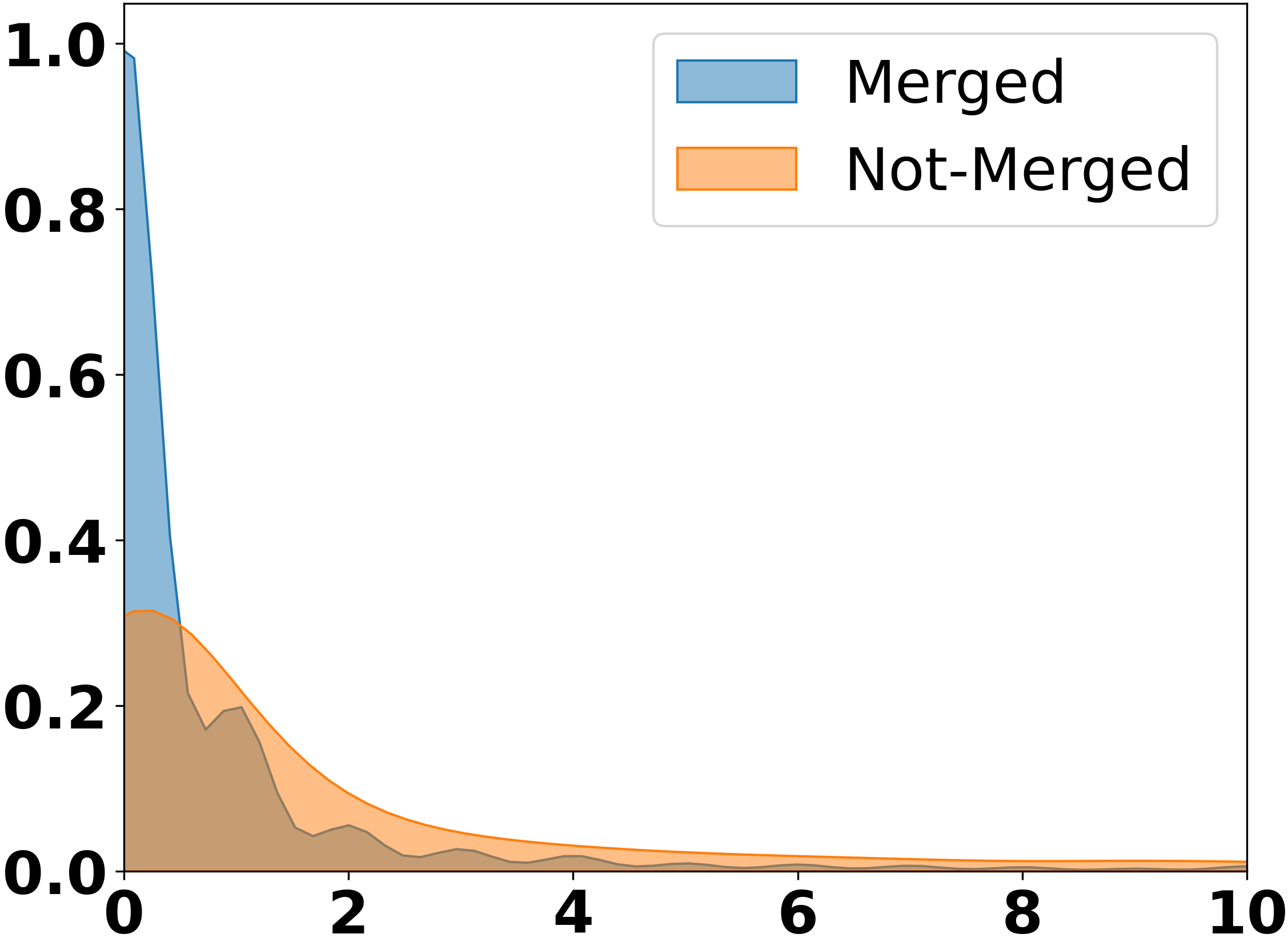}
        \caption{CI Fails}
        \label{fig:ci_fails}
    \end{subfigure}
    \caption{Differences in Merged vs. Not-merged PRs.}
    \label{fig:size_changes}
\end{figure}

Figures \ref{fig:file_changes} and \ref{fig:code_changes} show the differences in \textbf{code changes} between merged and not-merged PRs, in terms of \textit{\#LOC Changes} and \textit{\#File Changes}. Not merged PRs tend to introduce larger modifications in the number of files and LOC than merged ones. Based on Cliff’s $\delta$ (Table~\ref{tab:logit_effect}), the difference in total lines of code changes is 17\%, and the difference in the number of changed files is 10\%, both indicating a small-to-medium effect size that not-merged PRs skew toward larger changes. Figures \ref{fig:file_changes} and \ref{fig:code_changes} also show kernel density estimates of these two metrics. Because changes span several orders of magnitude, the distributions for these plots are shown on a \texttt{log10} scale, where each unit corresponds to a ten-fold increase in size. Higher density toward the right side of the plots reflects larger PRs. In both subfigures, the distribution for not-merged PRs is shifted slightly rightward relative to merged PRs, visualizing the same trend captured by Cliff’s $\delta$.

Figure \ref{fig:ci_fails} reports the distribution of \textbf{CI failures} for merged and not-merged agentic PRs. Not-merged PRs show a noticeably heavier tail, with many PRs accumulating multiple failing check runs. In contrast, merged PRs cluster sharply near zero, indicating that most merged contributions pass their checks with little or no failure. Cliff’s $\delta$ of 24\% indicates a moderate effect size, showing that not-merged PRs tend to experience more CI failures.

\begin{table}[h!]
\vspace{-0.3cm}
\centering
\caption{Logistic Regression and Effect Size (*$p$<0.05)}
\footnotesize
\label{tab:logit_effect}
\begin{tabular}{lrrrr}
\hline
\textbf{Characteristic} & \textbf{Coef} & \textbf{$p$-value} & \textbf{Odds Ratio} & \textbf{$\delta$} \\
\hline
\#LOC Changes     & -2.8e-06 & \textbf{$\sim$1\%$^{*}$} & 99\% & -0.17 \\
\#File Changes    & -0.0011  & \textbf{$\sim$1\%$^{*}$} & 99\% & -0.10 \\
\#Failed CI Checks & -0.1579  & \textbf{$\sim$1\%$^{*}$} & 85\% & -0.24 \\
\#Review Comments  & -0.0028  & $\sim$48\%          & 99\% & -0.05 \\
\#Review Revisions & -1.6e-05 & $\sim$67\%          & 99\% & -0.03 \\
\hline
\end{tabular}
\vspace{-0.3cm}
\end{table}

\begin{figure}[h]
    \vspace{-0.35cm}
    \centering
    \begin{subfigure}{0.49\columnwidth}
        \centering
        \includegraphics[width=0.85\textwidth]{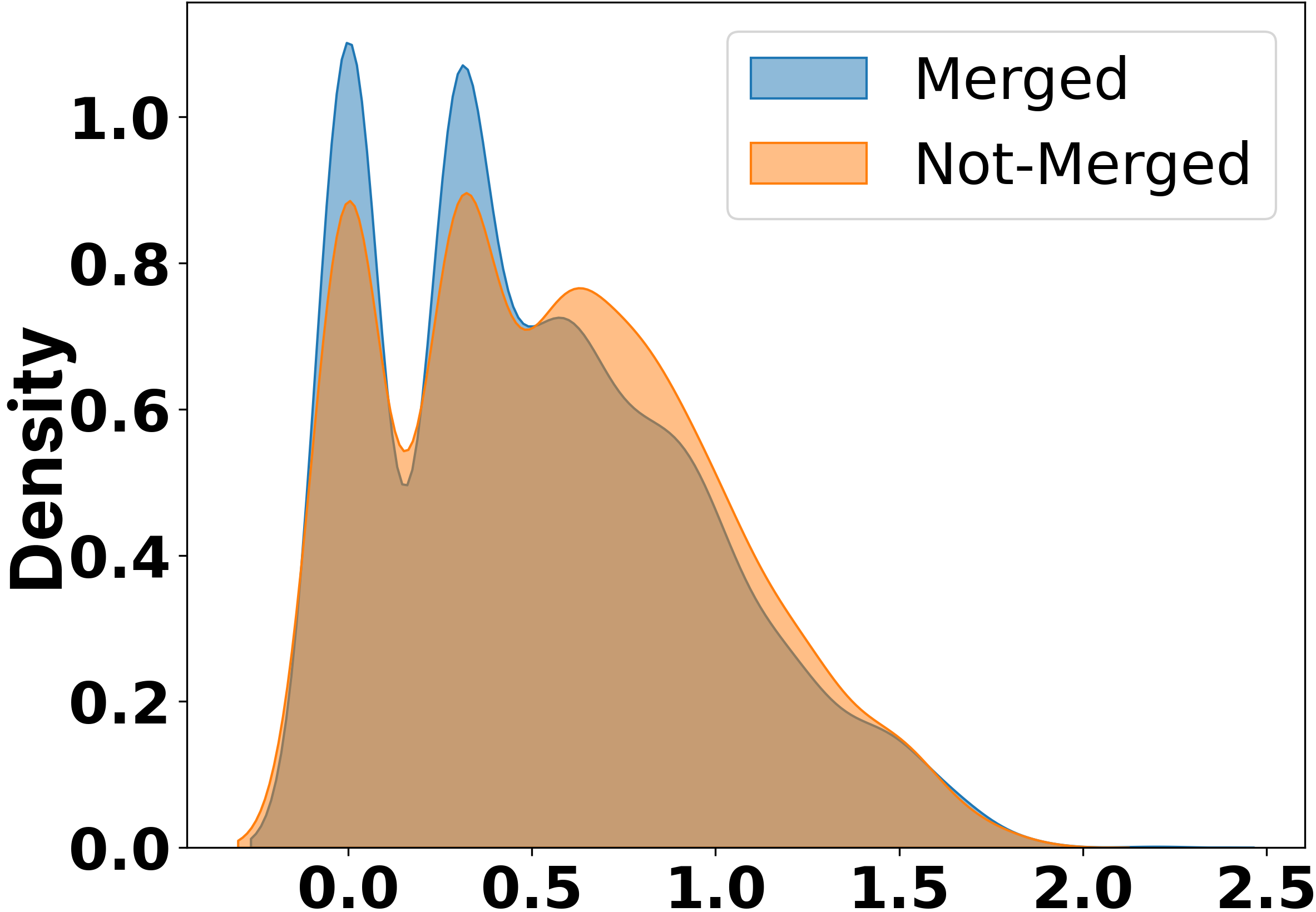}
        \vspace{-0.1cm}
        \caption{Comments in PRs (log10)}
        \label{fig:comments}
    \end{subfigure}
    \hfill
    \begin{subfigure}{0.49\columnwidth}
        \centering
        \includegraphics[width=0.82\textwidth]{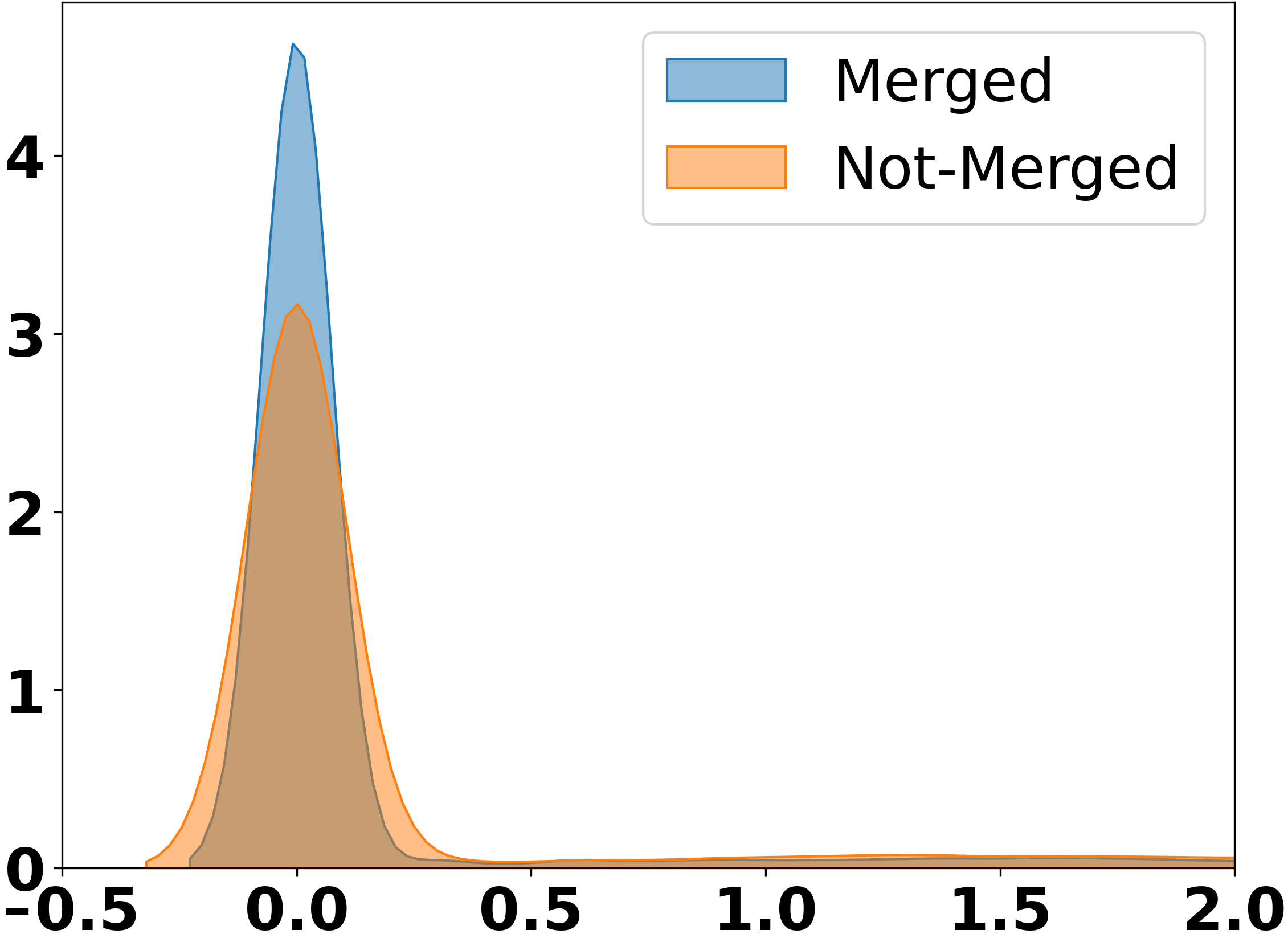}
        \vspace{-0.1cm}
        \caption{Revisions in PRs (log10)}
        \label{fig:revisions}
    \end{subfigure}
    \vspace{-0.1cm}
    \caption{Reviews of Merged vs. Not-merged PRs.}
    \label{fig:review_changes}
    \vspace{-0.4cm}
\end{figure}

Figure~\ref{fig:review_changes} shows the differences in \textbf{review interactions}. Not-merged PRs tend to receive more \textit{reviewer revisions} than merged ones. Based on Cliff’s $\delta$ (Table~\ref{tab:logit_effect}), developers make approximately 5\% more \textit{review comments} and 3\% more \textit{revisions} on not-merged PRs compared to merged PRs, although the effect sizes for both are small. The density distributions follow a similar overall pattern. However, the curves widen for not-merged PRs as the number of comments (Figure~\ref{fig:comments}) or revisions (Figure~\ref{fig:revisions}) increases, indicating that not-merged PRs often undergo extensive reviewer discussion and iterative refinement until a final decision is made to leave them open or mark them as rejected.

The logistic regression results in Table~\ref{tab:logit_effect} show that all model coefficients are negative, indicating that increases in these metrics are associated with lower odds of a PR being merged. Based on the $p$-values, all variables except \#\textit{review comments} and \#\textit{review revisions} are significant (threshold at $0.05$). Odds ratios quantify the practical effect of each metric. For example, a one-unit increase in \#\textit{LOC changes} decreases the odds of a merge by approximately 1\%, which can accumulate meaningfully given that code modifications span several orders of magnitude in size. Similar patterns hold for \#\textit{file changes}, \#\textit{review comments}, and \#\textit{revision cycles}, although with smaller magnitudes. Notably, each additional \textit{failed CI check} decreases the odds of a merge by about 15\%.

\vspace{-0.3cm}
\begin{tcolorbox}[colback=gray!25, breakable, colframe=gray!40, left=0pt, right=0pt, top=0pt, bottom=0pt]
Across all agentic PRs, \textit{documentation}, \textit{CI}, and \textit{build} tasks exhibit the highest success, while \textit{performance} and \textit{fix} see the lowest. Not-merged PRs tend to introduce larger code changes, touch more files, fail more CI checks, and receive slightly more review comments and revisions.
\end{tcolorbox}

\vspace{-0.1cm}
\noindent\textbf{RQ2: What patterns lead to agent-authored PRs not being merged in real-world software repositories?}

\noindent
Table~\ref{tab:failure_modes} summarizes the distribution of rejection patterns for the 600 manually annotated rejected agentic pull requests. We note that 38 PRs were no longer accessible at the time of analysis due to deletion or archival, leaving a total of 562 PRs for categorization.

\textsc{\textbf{Reviewer}}-level abandonment is the most frequent rejection pattern, accounting for 228 PRs (38\%). These PRs were left without any meaningful human reviewer interaction, often after prolonged inactivity or automated closure, indicating that a substantial fraction of agentic PRs fail before entering active review.

\textsc{\textbf{Pull request}}-level reasons form the second-largest group, comprising 188 PRs (31\%). Within this level, duplicate PRs are the most common pattern, affecting 142 PRs (23\%), where maintainers explicitly reference an existing PR that already implements the same change. For example, one PR was closed with the comment: \textit{``Superseded by PR \#715 which consolidates all GFQL code changes into a single PR"}~\cite{pr_link_8}. Unwanted features account for 24 PRs (4\%), where maintainers state that the contribution is misaligned with project goals or introduces excessive or unnecessary changes. Examples include: \textit{``Too old already superseded by more recent pushes}"~\cite{pr_link_9} and \textit{``This is a LOT to review, would really prefer smaller granular PRs"}~\cite{pr_link_7}.
Less frequent patterns include non-functional PRs (13; 2\%), which often consist of only setup or configuration tests. For example, PRs explicitly titled \textit{``testing DO NOT MERGE"}~\cite{pr_link_10}. Wrong task descriptions account for 7 PRs (1\%), where the PR description provides little to no meaningful context. Maintainers often respond with comments such as \textit{``Sorry, I don’t know what this is, but it doesn’t look like it belongs in our repo"}~\cite{pr_link_11}. Finally, wrong branch submissions (2; <1\%) occur when PRs are opened on incorrect branches, prompting maintainers to make comments such as \textit{``PR is opened against main. You probably want to open it against develop"}~\cite{pr_link_12}.

\textsc{\textbf{Code}}-level reasons represent the third most frequent category, affecting 133 PRs (22\%). The dominant pattern in this category is CI/test failure, observed in 99 PRs (17\%), where automated builds or tests fail due to the submitted changes. At times, reviewers even explicitly point out these failures, for example: \textit{``@copilot fix the merge conflicts; if you cannot fix these then close the PR"}~\cite{pr_link_6}.
Incorrect implementations (19; 3\%) and incomplete implementations (15; 2\%) comprise the remaining code-level failures. Reviewers often highlight technical inaccuracies or missing logic with comments such as: \textit{``The changes made to the billing.test.ts file are entirely wrong"}~\cite{pr_link_5}. % or \textit{``It still generates URLs with hyphens rather than the actual file path [Incomplete]"}~\cite{pr_link_4}.

\textsc{\textbf{Agentic}}-level issues are the least frequent category, comprising 13 PRs (2\%). Misalignment is the dominant pattern at this level, appearing in 9 PRs (1\%). In these cases, agents repeatedly fail to follow explicit reviewer instructions or misunderstand requested changes, even after multiple rounds of feedback. Reviewer comments often express frustration with comments such as \textit{``Devin stop being a dumb*ss, if you claim you "deleted 200 lines" then continue to"}~\cite{pr_link_3}. % or \textit{``Sorry Devin, this is a fail"}~\cite{pr_link_2}.
Licensing issues account for the remaining 4 PRs (<1\%). These PRs are rejected because agents do not comply with project-specific legal requirements, such as signing a Contributor License Agreement (CLA) or addressing ownership concerns. Maintainers explicitly reference these requirements with comments, e.g., \textit{``we ask that you sign our Contributor License Agreement before we can accept your contribution"}~\cite{pr_link_1}. Together, these examples highlight legal and governance constraints that current agents cannot satisfy.

\vspace{-0.2cm}
\begin{tcolorbox}[colback=gray!25, breakable, colframe=gray!40, left=0pt, right=0pt, top=0pt, bottom=0pt]
A majority of agentic PRs are not merged due to reviewer abandonment. Among reviewed PRs, the dominant rejection patterns include duplicate PRs, CI/test failures, and large or unwanted feature implementations.
\end{tcolorbox}
\vspace{-0.5cm}

\begin{table}[tb]
\centering
\small
\footnotesize
\caption{Taxonomy of Rejection Patterns in Agentic PRs.}
\resizebox{\columnwidth}{!}{%
\begin{tabular}{l p{1.6cm} p{5cm} p{0.5cm}}
\toprule
\textbf{Level} & \textbf{Pattern} & \textbf{Definition} & \textbf{Freq.} \\
\hline

\multirowcell{2}{\parbox{1cm}{\raggedright\textsc{Reviewer}}}
& Abandoned/Not Reviewed 
& PR closed with no meaningful human interaction; only bots (if any) performed actions. 
& 228 \\
\hline

\multirowcell{11}{\parbox{1cm}{\raggedright\textsc{Pull Request}}}

& Duplicate PR 
& Work already exists in another PR; maintainers explicitly reference the duplicate. 
& 142 \\
\cline{2-4}
& Unwanted Feature 
& Maintainers state that the feature is unnecessary, misaligned, or not intended.
% for the project. 
& 24 \\
\cline{2-4}
& Non-Functional PR 
& PR contains only setup, configuration, or scaffolding changes without functional contribution. 
& 13 \\
\cline{2-4}
& Wrong Task Description 
& PR description reflects misunderstanding of the task or intended change.
& 7 \\
\cline{2-4}
& Wrong Branch 
& PR targets the incorrect branch and must be resubmitted to the proper one. 
& 2 \\
\hline

\multirowcell{8}{\parbox{1cm}{\raggedright\textsc{Code}}}
& CI/Test Failure 
& PR fails automated build or deployment checks caused by its own changes. 
& 99 \\
\cline{2-4}
& Incorrect Implementation 
& Implementation is technically incorrect or solves the wrong problem despite a correct task description. 
& 19 \\
\cline{2-4}
& Incomplete Implementation 
& Contribution lacks required logic or completeness; reviewers flag missing or insufficient work. 
& 15 \\
\hline

\multirowcell{4}{\parbox{1cm}{\raggedright\textsc{Agentic}}}
& Misalignment 
& Agent fails to follow reviewer instructions or misunderstands explicit requested edits. 
& 9 \\
\cline{2-4}
& License Issues 
& Reviewers flag license or ownership concerns related to agent-generated content. 
& 4 \\

\bottomrule
\end{tabular}
}
\label{tab:failure_modes}
% \vspace{-0.87cm}
\end{table}
\section{Conclusion}
Our findings indicate that %the primary challenges faced by agentic PRs extend beyond code correctness. N
not-merged PRs tend to introduce larger and more invasive code changes, attempt broader feature additions, and exhibit higher rates of CI/test failures. Rejections of agentic PRs stem from multiple reasons, such as reviewer abandonment, duplicate PRs, or implementations of unwanted features. Overall, these results highlight difficulties of agents in task selection, coordination, and alignment with repository context. Maintainer feedback frequently emphasizes that PRs should be small, focused, and limited to a single coherent change, and discourages agentic submissions that combine substantive modifications with unrelated edits. The replication package for our study is publicly available~\cite{rep_pack}.%~\cite{pr_link_7}.

Improving the success of future agentic-AI workflows would require improving agents’ ability to identify existing or ongoing work, adhere to project contribution norms, decompose tasks into localized changes, and validate submissions against CI pipelines before opening new PRs. 
Failures of agentic PRs can also be socio-technical rather than purely technical. By characterizing the failure patterns of not-merged agentics PRs, our study provides empirical grounding for the design of more context-aware and collaboration-sensitive AI coding agents, and informs future research on integrating such agents into real-world software development workflows.

\balance
\bibliographystyle{ACM-Reference-Format}
\bibliography{ref}

\end{document}